\documentclass[aip,jcp,reprint,twocolumn]{revtex4-1}
\usepackage{graphicx}
\usepackage{color}
\usepackage[colorlinks=true, linkcolor=black, urlcolor=blue, citecolor=blue]{hyperref}

\begin{document}

\title{Solvation enhances folding cooperativity and the topology dependence
of folding rates in a lattice protein model}

\author{Nhung T. T. Nguyen}
\affiliation{Graduate University of Science and Technology, Vietnamese Academy
of Science and Technology, 18 Hoang Quoc Viet, Nghia Do, Cau Giay, Hanoi 11307, Vietnam}
\affiliation{Institute of Physics, Vietnamese Academy of Science and Technology, 10 Dao Tan, Ba Dinh, Hanoi 11108, Vietnam}

\author{Pham Nam Phong}
\affiliation{Faculty of Engineering Physics, Hanoi University of Science and
Technology, 1 Dai Co Viet Road, Hanoi, Vietnam}

\author{Duy Manh Le}
\affiliation{Laboratory of Advanced Materials and Natural Resources, Institute
for Advanced Study in Technology, Ton Duc Thang University, Ho Chi Minh City,
Vietnam}
\affiliation{Faculty of Applied Sciences, Ton Duc Thang University, Ho Chi Minh
City, Vietnam}

\author{Minh-Tien Tran}
\affiliation{Institute of Physics, Vietnamese Academy of Science and Technology, 10 Dao Tan, Ba Dinh, Hanoi 11108, Vietnam}

\author{Trinh Xuan Hoang}
\email[Corresponding author, Email: ]{txhoang@iop.vast.vn}
\affiliation{Institute of Physics, Vietnamese Academy of Science and Technology, 10 Dao Tan, Ba Dinh, Hanoi 11108, Vietnam}

\begin{abstract}
The aqueous solvent profoundly influences protein folding, yet its effects are
relatively poorly understood.  In this study, we investigate the impact of
solvation on the folding of lattice proteins by using Monte Carlo simulations.
The proteins are modelled as self-avoiding 27-mer chains on a cubic lattice,
with compact native states and structure-based G\=o potentials.  Each residue
that makes no contacts with other residues in a given protein conformation is
assigned a solvation energy $\varepsilon_s$, representing its full exposure to
the solvent. We find that a negative $\varepsilon_s$, indicating a favorable 
solvation, increases the cooperativity of the folding transition by lowering
the free energy of the
unfolded state, increasing the folding free energy barrier, and narrowing the
folding routes. This favorable solvation also significantly improves the
correlation between folding rates and the native topology, measured by the
relative contact order. Our results suggest that G\=o model may overestimate
the importance of native interactions and a solvation potential countering the
native bias can play a significant role. 
The solvation energy in our model can be related to the polar interaction
between water and peptide groups in the protein backbone.  It is therefore
suggested that the solvation of peptide groups may significantly contribute to
the exceptional folding cooperativity and the pronounced topology-dependence of
folding rates observed in two-state proteins. 
\end{abstract} 
\maketitle

\section{Introduction}

Water plays a fundamental role in protein folding, mainly through the dominance
of the hydrophobic effect \cite{Kauzmann1959,Dill1990,BaldwinRose2016} in the
folding and stabilization of proteins. Water also influences helix propensities
of amino acids \cite{Baldwin1999} and regulates protein aggregation
\cite{Thirumalai2012}. The extent of water-mediated effects on protein folding
remains elusive, despite extensive research. It has been shown experimentally
that solvation free energies are pairwise non-additive \cite{Kollman1995},
complicating the modeling and analysis of proteins by requiring many-body
interactions \cite{Papoian2004}. The particulate nature of water gives rise to
a desolvation barrier \cite{Pratt1977,Baker1997,Levy2006} to hydrophobic cluster
formation, which may contribute to the rate-limiting step in protein folding
\cite{Cheung2002,Chan2003jmb,Chan2005}. The role of solvents extends beyond
water, as changes in the solvent conditions such as pH, ionic strength, or the
presence of denaturants, can modulate the folding rates and outcomes
\cite{Jackson1991}.  Simulating folding kinetics at atomic resolution, whether
with explicit or implicit solvents, remains highly computationally demanding
\cite{Pande2004,Shaw2011}. Meanwhile, valuable insights into the role of
solvents on protein folding can be obtained using simple lattice
\cite{Hao1997,Sorenson1998}, off-lattice \cite{Cheung2002,Chan2003jmb,Chan2005}
and continuous \cite{Kamien2005,Banavar2007} models.

In this study, we examine the role of solvation in determining two remarkable
properties of two-state proteins---a class of small, single-domain proteins
that fold essentially with simple two-state kinetics
\cite{Serrano1994,Plaxco_SH3}. One is folding cooperativity, which refers to
the ability of proteins to exhibit a sharp transition between the unfolded and
folded states without significant intermediate states, resembling an
``all-or-none'' process \cite{Chan2004}.  Two-state proteins exhibit
surprisingly high folding cooperativity in both kinetic and thermodynamic
aspects, compared to those obtained from simulations
\cite{Kaya2000prl,Kaya2000prot,Kaya2003prl}.  The other property is the
topology dependence of folding rates, an important observation made by Plaxco
et al. \cite{Plaxco1998,Plaxco2000}. These authors discovered a strong
correlation between the logarithms of experimentally observed folding rates for
two-state proteins and the relative contact order \cite{Plaxco1998}, which is
defined as the average sequence separation of native contacts relative to the
protein chain length. This empirical result called into question early theories
and simulations regarding the folding rates, many of which suggested that chain
length is a dominant factor
\cite{Thirumalai,Shakhnovich,Cieplak_prl,Cieplak2000,Cieplak2001,FinBadredtinov}.
Despite significant progress
\cite{Debe1999,Eaton1999,Dill2003,Makarov2003,Fersht2000,Dobson2005,Plaxco2003,Kaya2003prot,Chan2005topomer,Chan2008,Chan2009,Chan2011,Chan2013,KogaTakada,Cieplak_bj,Cieplak2004},
our understanding of the physical basis for the rate-topology dependence
remains limited.  It was suggested that the rate-topology dependence is a
consequence of the extraordinary folding cooperativity in two-state proteins
\cite{Plaxco2003}.  Thus, our goal is to check if solvation contributes to the
folding cooperativity and whether its increase is associated with the increase
in the rate-topology dependence.

G\=o \cite{Go1975} and G\=o-like models \cite{Hoangjcp1,Hoangjcp2,Clementi} have
been widely used for studying the folding mechanism
\cite{Baker2000,Takada2019}, offering insights into how the folding process is
driven solely by native interactions. Despite this global native preference,
these models show moderate folding cooperativity
\cite{Kaya2000prl,Kaya2000prot,Chan2004} and insignificant to low
correlations between folding rates and the relative contact order
\cite{KogaTakada,Cieplak_bj,Cieplak2004}. Jewett et al. \cite{Plaxco2003}
demonstrated that adding an explicit energetically cooperative feature to the
Hamiltonian of the lattice G\=o model can give rise to the rate-topology
correlation. Another modification to the G\=o model that leads to a similar
effect was given by Kaya and Chan \cite{Kaya2003prot}, who introduced a
coupling between non-local contact interactions and local conformational
preferences.  These studies indicate that an energetic component leading to
folding cooperativity is missing in the G\=o model. More recently, by using an
off-lattice G\=o-like model with desolvation barriers in the pairwise
potentials, Chan and coworkers \cite{Chan2008,Chan2009,Chan2011,Chan2013}
showed that the desolvation barriers can substantially enhance folding
cooperativity and the diversity in the folding rates, while their impact on
the rate-topology correlation is modest. 

Here, we propose a modification to the G\=o model by incorporating a solvation
energy term that favors fully solvent-exposed residues. This modification is
based on a hypothesis that G\=o potentials may overemphasize the contributions
of native interactions to folding free energy, thereby undermining favorable
solvent-protein interactions in the unfolded state, such as the polar
interaction between water and exposed peptide groups in an unfolded protein. 
Not only competing with native interactions, the proposed solvation energy can
also energetically disfavor the formation of non-native contacts. Finally, the
solvation energy term represents a many-body interaction, enabling it to
influence a wide range of conformations in a
pairwise non-additive manner. We will show that this modification produces
intriguing effects on folding cooperativity and the rate-topology dependence of
the protein model.

\section{Model and Methods}

\begin{figure}
\includegraphics[width=3.4in]{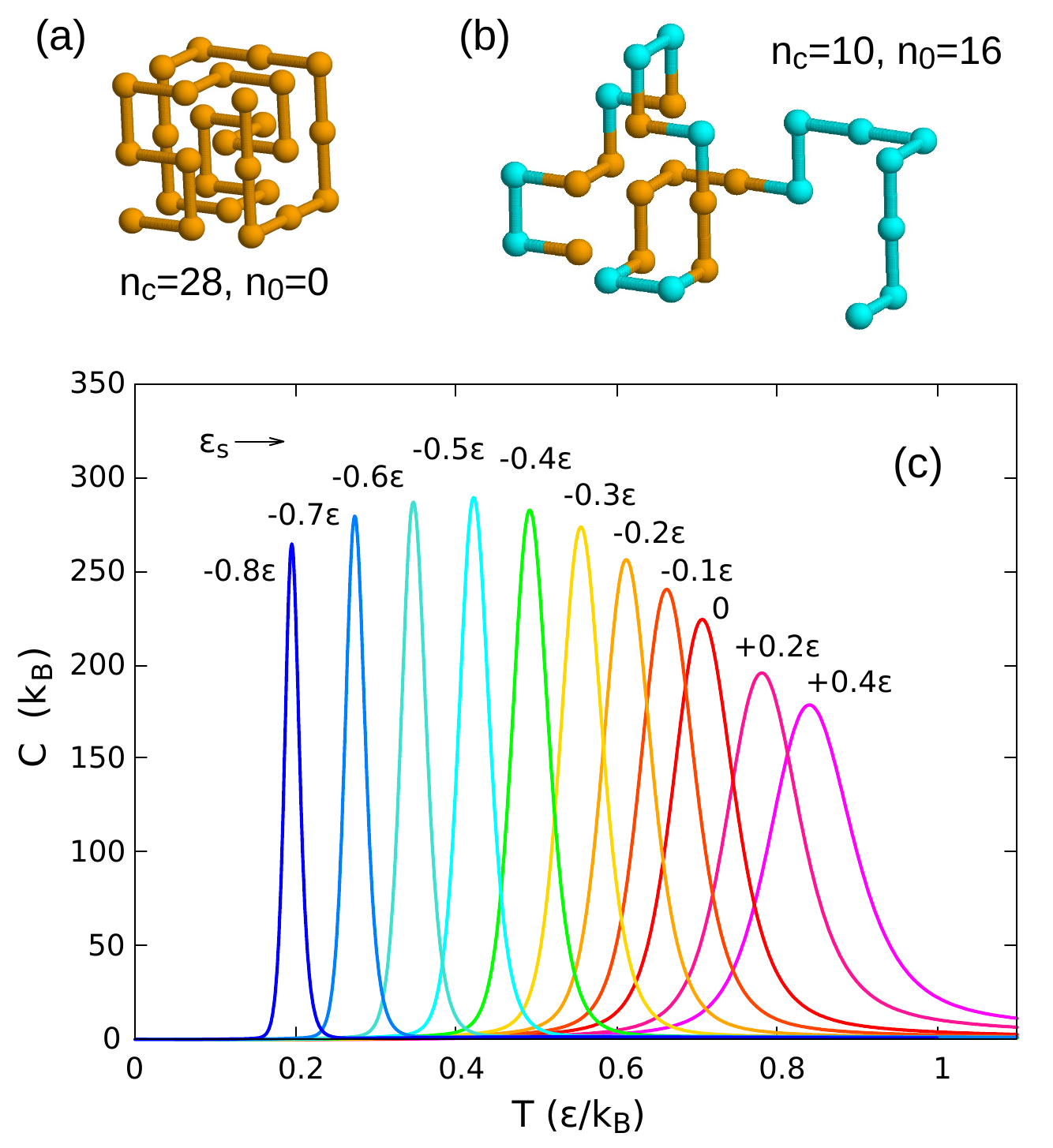}
\caption{Examples of lattice protein conformations of 27 residues: (a) a
compact native state and (b) a partially unfolded conformation.  In these
visualizations, residues in contact with other residues are shown in orange,
while residues fully exposed to the solvent are shown in cyan.  The number of
native contacts ($n_c$) and the number of fully exposed residues ($n_0$) for
each conformation are indicated.  (c) Temperature dependence of the specific
heat for the model protein with the native state shown in (a), for various
values of the solvation energy parameter $\varepsilon_s$ ranging between
$-0.8\varepsilon$ and $+0.4\varepsilon$, as indicated. 
} 
\label{fig1}
\end{figure}

We model proteins as self-avoiding chains on a three-dimensional cubic lattice
with G\=o potentials \cite{Go1975} for contacts between the residues and an
implicit solvent.  The proteins are of the same length of 27 residues and have
the native states being $3\times 3 \times 3$ compact conformations (see an
example of a native conformation in Fig. \ref{fig1}(a)). 
The energy of a protein in a
given conformation is given by
\begin{equation}
E = -n_c \varepsilon + n_0 \varepsilon_s ,
\label{eq1}
\end{equation} 
where $n_c$ is the number of native contacts, $\varepsilon > 0$ is an energy
unit in the system with $-\varepsilon$ being the energy of a native contact;
$n_0$ is the number of residues that make no contacts with other residues,
considered as the number of fully solvent-exposed residues (see example in Fig.
\ref{fig1} (b)); and $\varepsilon_s$ is the solvation energy per fully
solvent-exposed residue. In a given protein conformation $n_c$ is given by
\begin{equation}
n_c = \sum_{i<j-1} \Delta_{ij} \Delta^\mathrm{NAT}_{ij} ,
\label{eq2}
\end{equation} 
where $\Delta$ is a protein contact map in the given conformation with
$\Delta_{ij}$ equal to 1 if two residues $i$ and $j$ are nearest neighbors on
the lattice but not consecutive along the chain, and $0$ otherwise;
$\Delta^\mathrm{NAT}$ is the contact map of the native conformation. 
We generally consider $\varepsilon_s \leq 0$, but also some positive values of
$\varepsilon_s$ for comparison.  For $\varepsilon_s=0$, the model becomes
identical to the G\=o model \cite{Go}. 

The relative contact order parameter (RCO) \cite{Plaxco1998} of a native state 
is defined as 
\begin{equation}
RCO = \frac{\sum_{i<j-1} \Delta_{ij}^{NAT} |i-j|}{N\sum_{i<j-1} 
\Delta_{ij}^{NAT}} \;\;.
\end{equation}
There are 97 distinct values of RCO ranging from about 0.28 to about 0.53 for
$3 \times 3 \times 3$ compact conformations \cite{Plaxco2003}.  Our set of
proteins corresponds to 97 different native states, such that for each RCO
value there is only one native state picked up randomly from the compact
conformations. The number of native contacts is $n_c=28$ for all maximally
compact lattice proteins.

Our Monte Carlo (MC) simulations follow the Metropolis algorithm
\cite{Metropolis1953} and the standard polymer move set, which contains single
monomer moves (corner flip and end move) and double monomer moves (crankshaft)
\cite{Henkel}.  The single and double monomer moves were attempted with
probabilities of 0.2 and 0.8, respectively. For calculating thermodynamic
properties, we employed the replica-exchange technique \cite{Swendsen1986} in
parallel tempering simulations and the weighted histogram method
\cite{Ferrenberg} in the data analyses. The specific heat of a protein is
given by
\begin{equation}
C(T) = \frac{\langle E^2 \rangle - \langle E \rangle^2}{k_B T^2} ,
\end{equation} 
where $\langle \cdot \rangle$ denotes a thermodynamic average, $k_B$ is
the Boltzmann constant, and $T$ is an absolute temperature. In our
consideration, $T$ is given in units of $\varepsilon/k_B$.

Kinetic properties were obtained from independent folding and unfolding
trajectories in multiple simulations. The folding trajectories start from 
random conformations sampled at infinite temperatures
and proceed until the native state is reached.  The
unfolding trajectories start from the native state and proceed until the number
of native contacts is less than 20\% of that of the native state. The folding
and unfolding rates ($k_f$ and $k_u$) are determined as the inverse of the
median folding and unfolding times, respectively. Time is measured in 
MC steps.

\section{Results}

We began by studying the effects of solvation on the folding properties of
proteins. For a detailed analysis, we selected a representative protein with an
intermediate RCO value. The native state of this protein is illustrated in Fig.
\ref{fig1}(a), and its corresponding RCO value is 0.4048.  We calculated the
temperature dependence of the specific heat for this protein at various values
of $\varepsilon_s$, between $-0.8\varepsilon$ and $+0.4\varepsilon$.  Fig.
\ref{fig1}(c) shows that as $\varepsilon_s$ decreases, the peak of the specific
heat shifts to lower temperatures and becomes sharper.  The increased sharpness
of the peak with decreasing $\varepsilon_s$ is accompanied by a reduced width
and an increased height of the peak; however, the height reaches its maximum at
$\varepsilon_s \approx -0.5\varepsilon$ and then decreases. These remarkable
changes in the specific heat peak indicate that solvation significantly
influences the folding transition.

\begin{figure}
\includegraphics[width=3.4in]{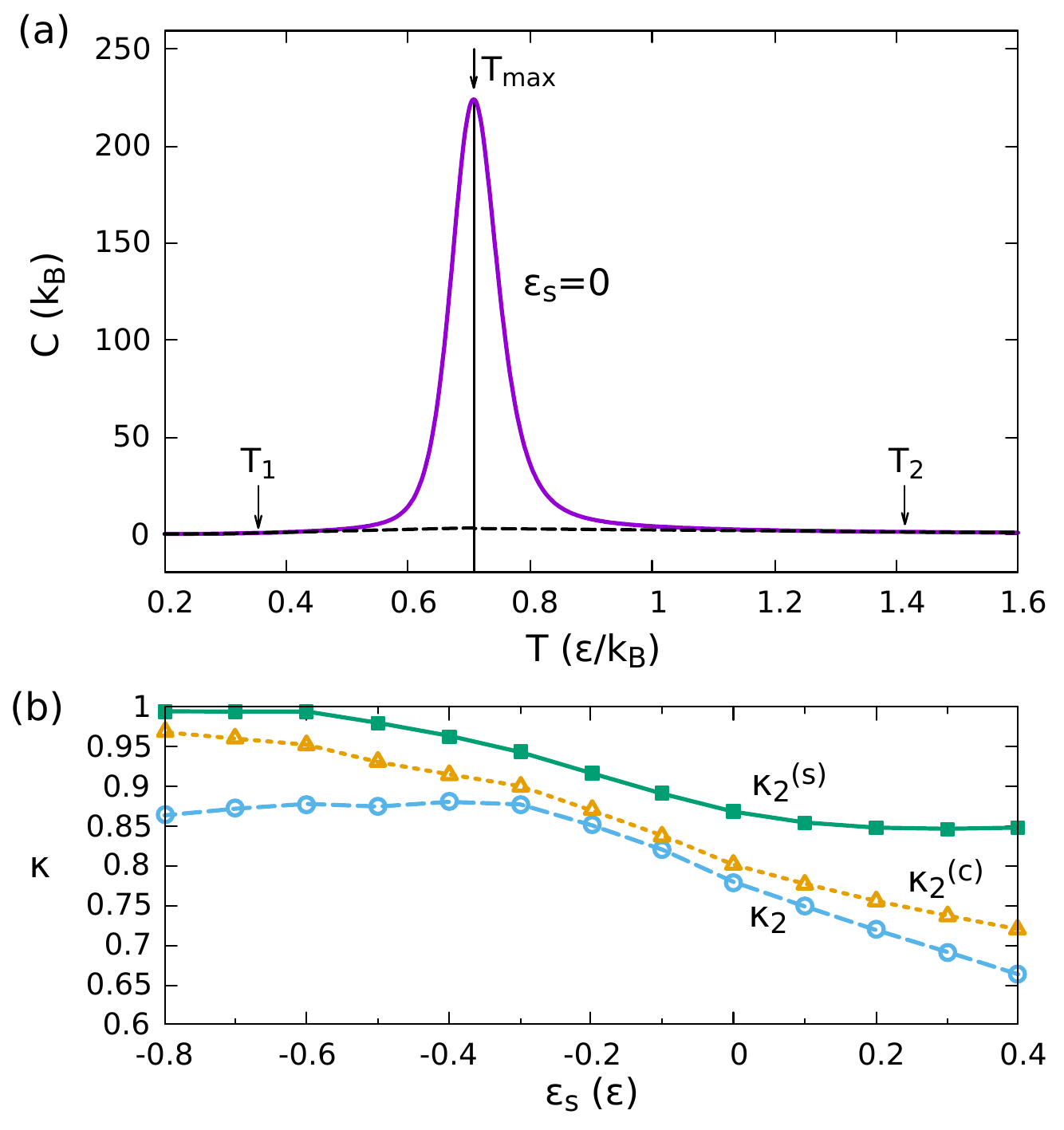}
\caption{
(a) The specific heat as a function of temperature, $C(T)$, 
for the model protein in Fig. \ref{fig1} with $\epsilon_s=0$,
is shown with baselines (dashed) for the folded and unfolded states.
The latter are defined as straight lines
tangent to the curve $C(T)$ at a low temperature $T_1$ and a high temperature
$T_2$, respectively, and extending between these temperatures and
the temperature of the specific heat maximum,
$T_\mathrm{max}$. For $T<T_1$ and $T>T_2$, the baseline is the same as $C(T)$.
For convenience, we have chosen $T_1 = 0.5\, T_\mathrm{max}$ and $T_2=2\,
T_\mathrm{max}$. 
(b) Dependence of the cooperativity index $\kappa$, calculated
with a baseline subtraction ($\kappa_2^{(s)}$ (squares)) and without
baseline subtraction ($\kappa_2$ (circles) and $\kappa_2^{(c)}$
(triangles)) from the specific heat, on the solvation energy parameter
$\varepsilon_s$, for the same protein as in (a).
}
\label{fig2}
\end{figure}

The cooperativity of the folding transition can be assessed from the specific
heat data. We followed Privalov and Potekhin \cite{Privalov1986},
and Kaya and Chan \cite{Kaya2000prl,Kaya2000prot}, to calculate the ratio
$\kappa=\Delta H_\mathrm{vh}/\Delta H_\mathrm{cal}$ between the van't Hoff
enthalpy ($\Delta H_\mathrm{vh}$) and the calorimetric enthalpy ($\Delta
H_\mathrm{cal}$) from the specific heat. 
The experimental calorimetric criterion requires $\kappa=1$
for a two-state process \cite{Privalov1986}. Two-state proteins have $\kappa$
within the range of $1 \pm 0.05$ \cite{Privalov1986,Serrano1994}. Though the
calorimetric criterion itself is not sufficient for 
determining a two-state process \cite{Zhou1999}, $\kappa$ may be considered as
a measure of thermodynamic cooperativity \cite{Kaya2000prl,Kaya2000prot}.
Without a baseline subtraction from the specific heat, $\kappa$ can be
given by \cite{Privalov1986,Kaya2000prot} 
$\kappa_2 = 2 T_\mathrm{max} \sqrt{k_B
C_\mathrm{max}}/\Delta H_\mathrm{cal}$ with
$\Delta H_\mathrm{cal} =\int_0^\infty C(T) dT$,
where $C_\mathrm{max}$ and $T_\mathrm{max}$ are the specific heat maximum
and its temperature. In our numerical calculation of $\Delta
H_\mathrm{cal}$, the integration of $C(T)$ is taken with $T$ running from 0 to
$5T_\mathrm{max}$.
When a baseline subtraction is applied, $\kappa$ is equal to $\kappa_2^{(s)}$,
as given in Ref. \cite{Kaya2000prot} under the same notation.
Our construction of the baselines (see Fig. \ref{fig2}(a) and
its caption) is similar to that described in Ref. \cite{Kaya2000prot}, with
an addition that we specified the two temperatures, where the baselines for the
folded and unfolded states are tangent to the
specific heat curve, to be given by $T_1=0.5 T_\mathrm{max}$ and
$T_2=2 T_\mathrm{max}$, respectively.
We defined also another cooperativity index, denoted as $\kappa_2^{(c)}$,
which is without baseline subtraction but with 
$\Delta H_\mathrm{cal} =\int_{T_1}^{T_2} C(T) dT$.  
By comparing $\kappa_2^{(c)}$ to $\kappa_2$ and $\kappa_2^{(s)}$, one can see
separately the effects of the temperature range and of the baselines,
respectively, on their values. The numerical values of
$\kappa_2^{(c)}$ and $\kappa_2^{(s)}$ depend on the choice of $T_1$ and $T_2$,
and it is not clear how to determine the correct values. Our aim here is to
show how they qualitatively depend on $\varepsilon_s$.

Figure \ref{fig2}(b) shows that as $\varepsilon_s$ decreases from
$0.4\varepsilon$ to $-0.8\varepsilon$, $\kappa_2^{(s)}$ monotonically increases
from approximately 0.85 to more than 0.99.
$\kappa_2^{(c)}$ also monotonically increases to a value larger than
0.96, while $\kappa_2$ only increases
from 0.66 to a maximum of about 0.88 at $\varepsilon_s=-0.4\varepsilon$ and
then slightly decreases. The main reason why $\kappa_2$ decreases is that
at low $\varepsilon_s$, the specific heat develops a long
tail at high temperatures, leading to an increase in the calculated
calorimetric enthalpy. Therefore, 
$\kappa_2^{(s)}$ and $\kappa_2^{(c)}$ may be more meaningful than
$\kappa_2$ in underscoring the folding transition for low values of
$\varepsilon_s$. Both the behaviors of
$\kappa_2^{(s)}$ and $\kappa_2^{(c)}$ indicate a significant increase of folding
cooperativity upon decreasing the solvation energy parameter.

\begin{figure}
\includegraphics[width=3.4in]{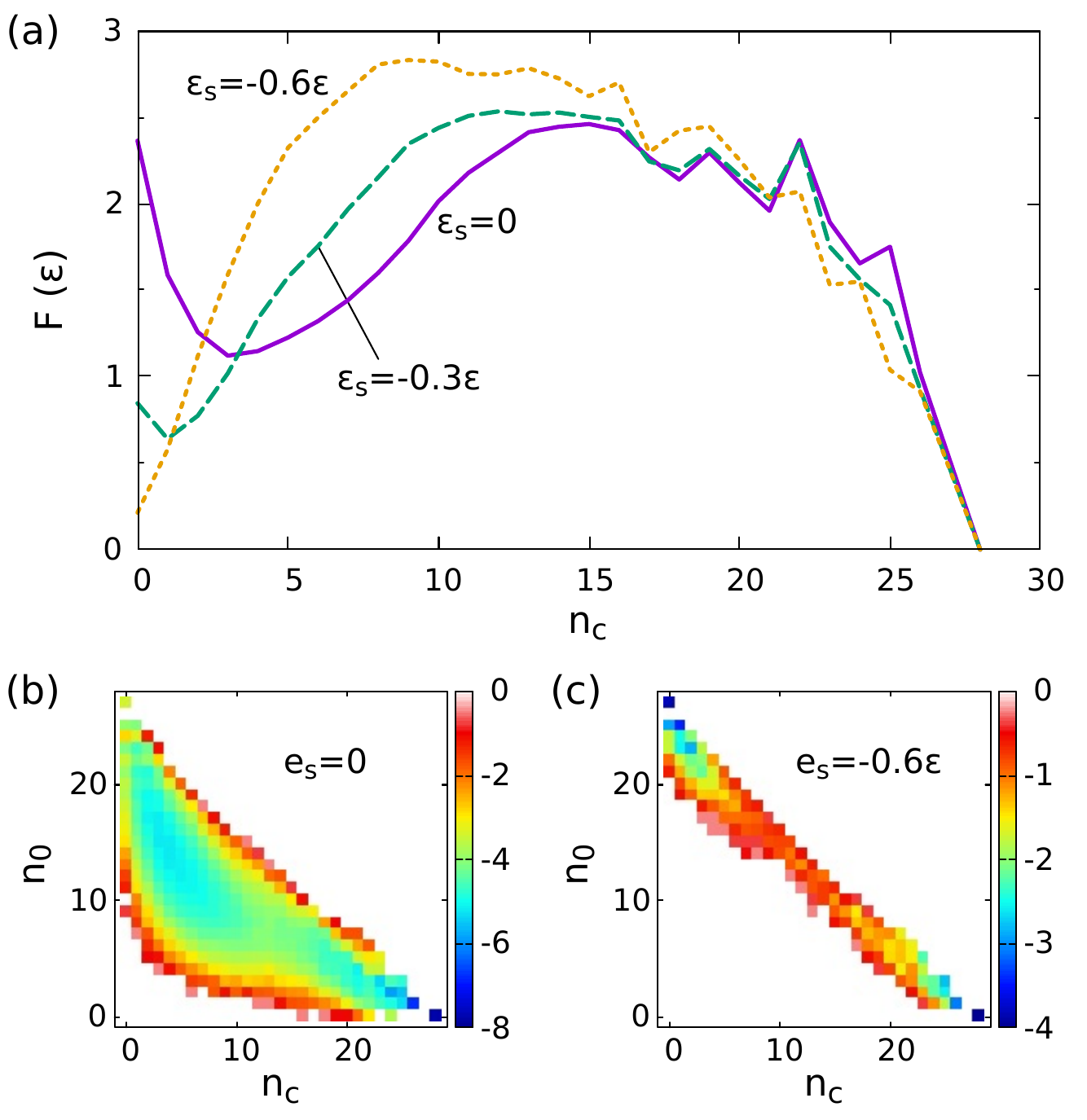}
\caption{
(a) The free energy $F$ as a function of the number of native contacts $n_c$ at
the temperature $T=T_\mathrm{max}$ for the protein with the native state shown
in Fig. \ref{fig1}(a), with the solvation energy parameter $\varepsilon_s=0$
(solid), $\varepsilon_s=-0.3\varepsilon$ (dashed) and
$\varepsilon_s=-0.6\varepsilon$ (dotted). $F$ is calculated as $F(n_c)=-k_B
T_\mathrm{max} \log P(n_c)$, where $P(n_c)$ is the probability of finding a
conformation with $n_c$ native contacts, obtained from a long MC simulation at
$T=T_\mathrm{max}$.  (b and c) The free energy as a function of $n_c$ and the
number of fully solvent-exposed residues $n_0$, shown as color maps for the two
cases: $\varepsilon_s=0$ (b) and $\varepsilon_s=-0.6\varepsilon$ (c).  The maps
were calculated using the same simulation data as in (a).  The native state
corresponds to the state with $n_c=28$ and $n_0=0$.
}
\label{fig3}
\end{figure}

The effect of solvation on folding cooperativity can be understood by examining
the free energy profiles of the protein (Fig. \ref{fig3}). These profiles show
that a negative solvation energy shifts the unfolded state and the transition
state further away from the native state along the folding coordinate,
increases the folding free energy barrier, and narrows the folding pathways.
Specifically, Fig. \ref{fig3}(a) shows that while the native state remains
unchanged, the unfolded state, corresponding to the free energy minimum at a
small number of native contacts $n_c$, is shifted toward lower values of $n_c$
and also lower values of the free energy $F$ as $\varepsilon_s$ decreases from
zero to $-0.6\varepsilon$.  At the same time, the transition state, which
presumably corresponds to the maximum of the largest free energy barrier, is
shifted toward lower values of $n_c$ and higher values of $F$.  Due to these
shiftings, the folding free energy barrier, $\Delta F_{U}^{\ddag}$, given by
the free energy difference between the transition state ($\ddag$) and the
unfolded state ($U$), obtained at $T=T_\mathrm{max}$, is increased almost twice
(from $\sim$1.34$\varepsilon$ to $\sim$2.62$\varepsilon$) as $\varepsilon_s$
decreases from zero to $-0.6\varepsilon$. Since $T_\mathrm{max}$ also decreases
with $\varepsilon_s$ (see Fig. \ref{fig1}), the increase of $\Delta
F_{U}^{\ddag}$ in units of $k_B T$ is about 4 times as $\varepsilon_s$
decreases from zero to $-0.6\varepsilon$.  Figures \ref{fig3} (b and c) show
that the pathway between the native state and the unfolded state is narrowed
due to the effect of a negative solvation energy, as the latter energetically
disfavors conformations with low values of $n_0$, the number of fully
solvent-exposed residues. Because non-native contacts have zero energy, they
are strongly disfavored by the negative solvation energy if their formation
decreases $n_0$.  Both the increased free energy barrier and the narrowed
folding pathway promote folding cooperativity, as they make the accessible
states on the pathway connecting the unfolded and native states less probable,
making the system more two-state-like.
We have checked that the linear shape
of folding pathway in the $n_0$-$n_c$ plan shown in Fig. \ref{fig3}c 
is also found in other lattice proteins for the same value of $\varepsilon_s$.

\begin{figure}
\includegraphics[width=3.4in]{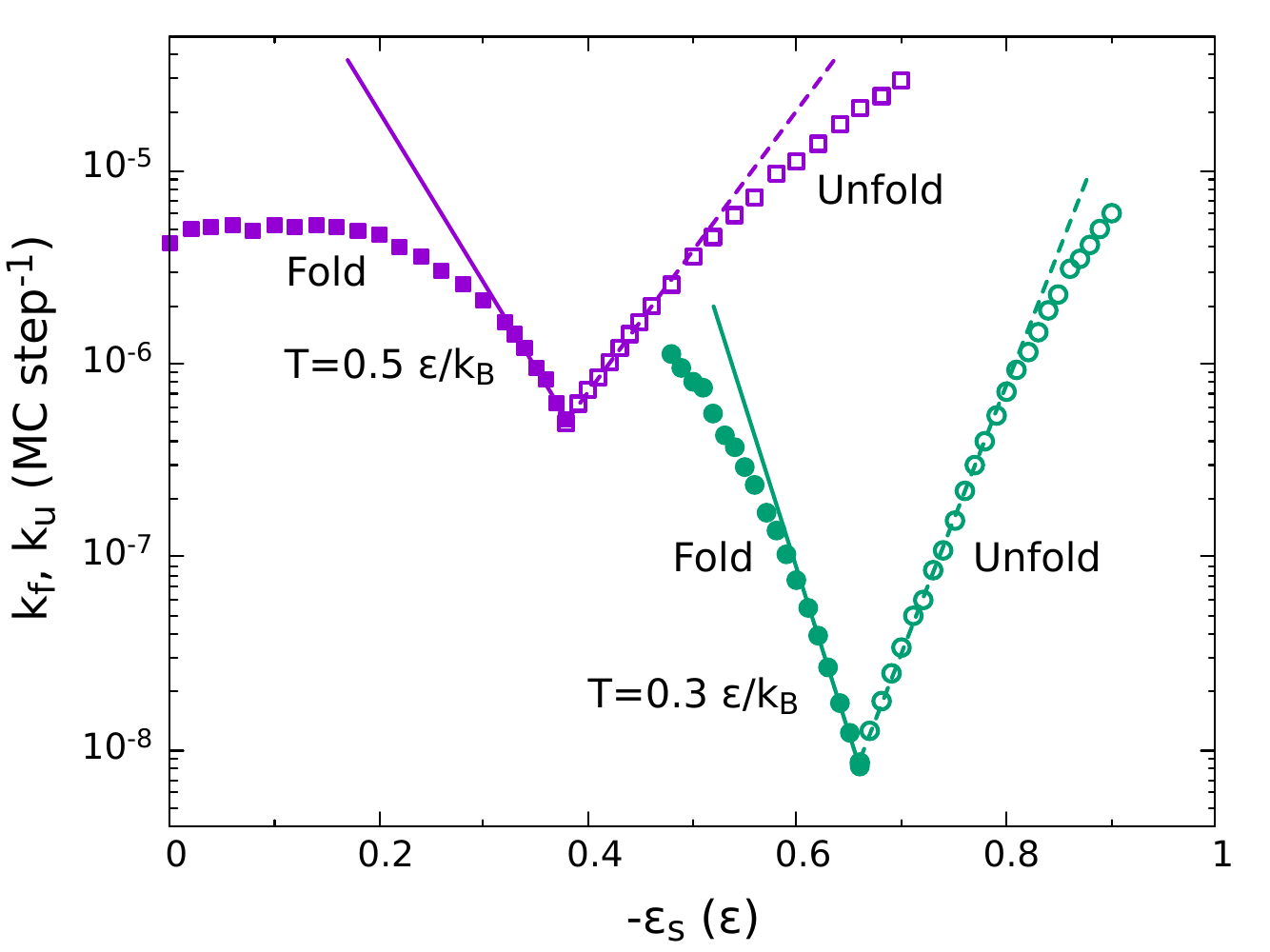}
\caption{Chevron plots of the folding (filled symbols) and unfolding (open
symbols) rates as functions of the solvation energy parameter $\varepsilon_s$
for the protein with the native state shown in Fig. \ref{fig1}(a) at two
different temperatures, $T=0.5~\varepsilon/k_B$ (squares) and
$T=0.3~\varepsilon/k_B$ (circles).  The folding ($k_f$) and unfolding ($k_u$)
rates, shown on a logarithmic scale, are defined as the inverse of the median
folding (unfolding) time obtained from the simulations.  For each temperature
and each value of $\varepsilon_s$, $k_f$ and $k_u$ are determined from 1001
independent trajectories.  Data points near the chevron turnover are fitted by
linear functions for the folding (solid line) and unfolding (dashed line) rates
separately.
}
\label{fig4}
\end{figure}

\begin{figure}
\includegraphics[width=3.4in]{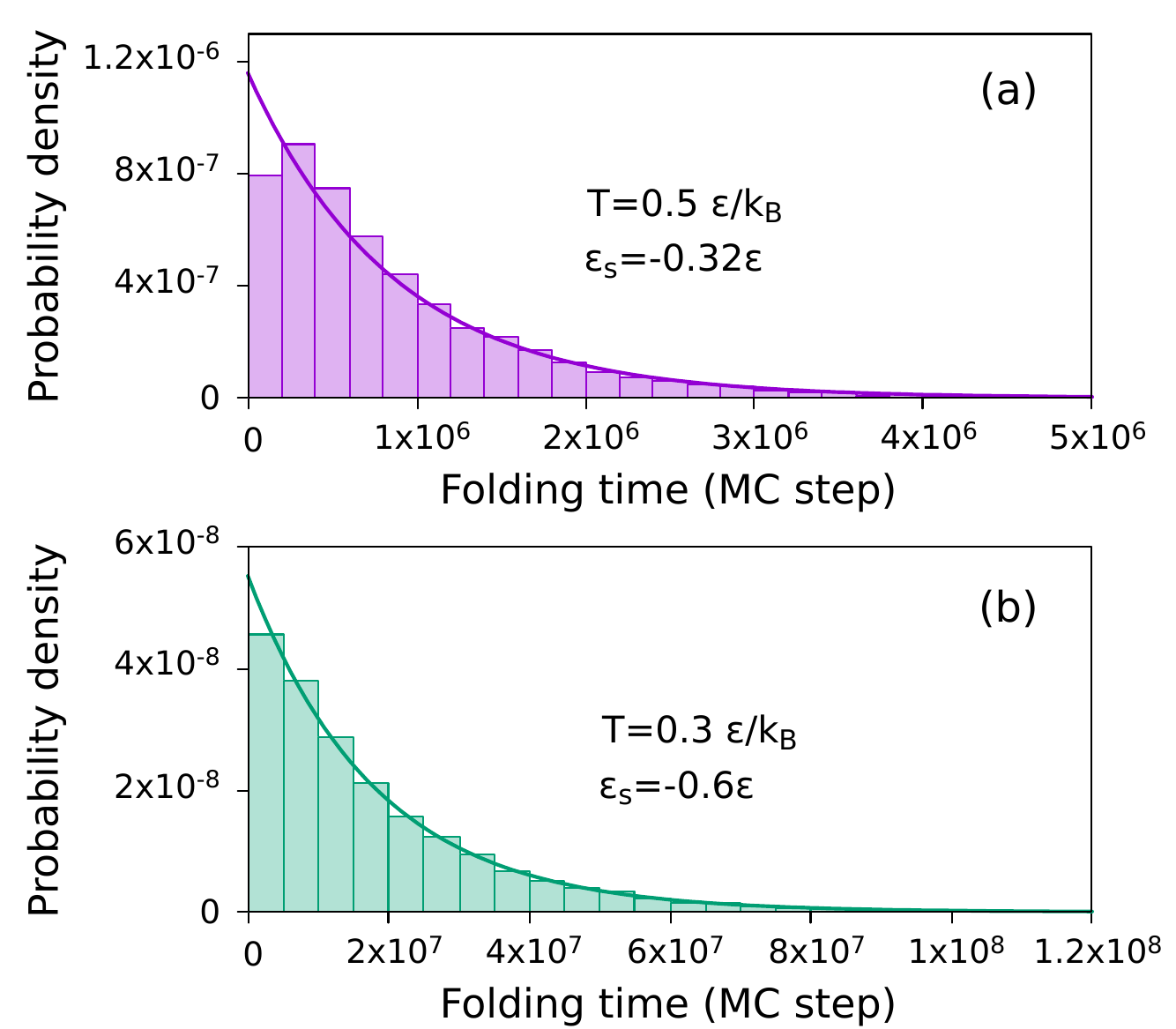}
\caption{Normalized histograms of the folding times (boxes) for the
protein considered in Fig. \ref{fig4} at two different conditions of
temperature and solvation energy: (a) $T=0.5~\varepsilon/k_B$,
$\varepsilon_s=-0.32\varepsilon$; and (b) $T=0.3~\varepsilon/k_B$,
$\varepsilon_s=-0.6\varepsilon$.  The folding times in each plot are obtained
from simulations of 10,000 independent folding trajectories.  The histograms
are plotted against the exponential distribution $g(t)=t_a^{-1} \exp(-t/t_a)$
(solid line) of the folding time $t$, where $t_a$ is the average folding
time calculated from the simulation data.} 
\label{fig5}
\end{figure}

One can think of varying the solvation energy parameter as a way to mimic the
effect of denaturants.  We determined the folding and unfolding rates, $k_f$
and $k_u$, at two different temperatures, $T=0.5~\varepsilon/k_B$ and
$T=0.3~\varepsilon/k_B$, from the simulations of the selected protein for
various values of $\varepsilon_s$. Using these data, we constructed chevron
plots of $\log k_f$ and $\log k_u$ vs.  $-\varepsilon_s$, where
$-\varepsilon_s$ represents the effect of increasing denaturant concentration
(Fig. \ref{fig4}).  Note that the pure solvent may correspond to a specific
value of $\varepsilon_s$, and the presence of a denaturant adds to that value.
Figure \ref{fig4} shows that as the temperature decreases, the V-shaped region
around the denaturation midpoint shifts toward higher values of $-\varepsilon_s$
and lower values of $k_f$ and $k_u$. The linear fits of $\log k_f$ and $\log
k_u$ vs. $-\varepsilon_s$ near the denaturation midpoint are indications of
two-state folding kinetics.  Note that these linear fits are better defined,
spanning a wider range of $-\varepsilon_s$, at $T=0.3~\varepsilon/k_B$ than at
$T=0.5~\varepsilon/k_B$. This indicates that the kinetic data align more
closely with the two-state model in the lower range of $\varepsilon_s$.
Experimentally, the movements of the denaturation
midpoint towards higher denaturant concentration and lower values of 
transition rates on decreasing temperature have been observed, e.g. for 
the N-terminal domain of L9 protein \cite{Raleigh1998,Raleigh2008}. 
For this protein, no rollover of the chevron plots was observed, even at
low concentrations of denaturant, for temperatures between 9$^\circ$C and
40$^\circ$C \cite{Raleigh1998,Raleigh2008}, indicating 
a very high two-state kinetic cooperativity that remains unchanged
within this range of temperature. 
For temperatures above 55$^\circ$C, the range of denaturant
concentration at which the rates can be measured is significantly reduced
\cite{Raleigh1998}, which somewhat aligns with our simulation results.

Figure \ref{fig5} shows that the histograms of the folding times obtained
from simulations for the temperature and solvation energy parameters, $T$
and $\varepsilon_s$, within the linear regions of the corresponding
chevron plots in Fig. \ref{fig4}, align well with an exponential
distribution of folding time, which is characteristic of the kinetics of a
two-state system \cite{Plaxco_SH3}. Notably, the histogram at
$T=0.3~\varepsilon/k_B$ and
$\varepsilon_s=-0.6\varepsilon$ (Fig.~\ref{fig5}b) fits the exponential
function slightly better at low time
values than the one at $T=0.5~\varepsilon/k_B$ and
$\varepsilon_s=-0.32\varepsilon$ (Fig.~\ref{fig5}a), indicating 
an increased two-state cooperativity at the lower $\varepsilon_s$ value.

\begin{figure}
\includegraphics[width=3.4in]{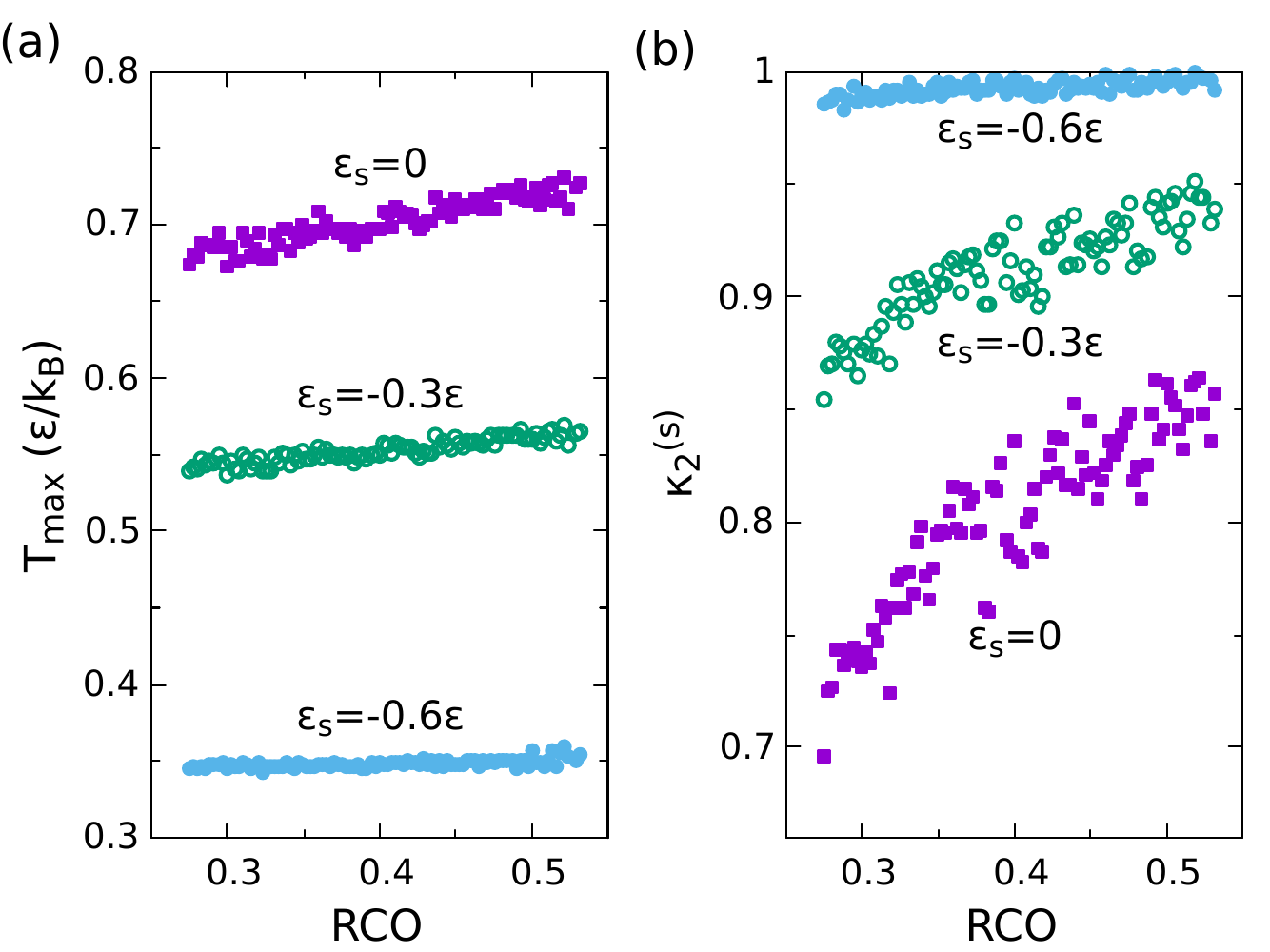}
\caption{Temperature of the maximum of the specific heat, $T_\mathrm{max}$, (a)
and the cooperativity index $\kappa_2^{(s)}$ (b) of 97
lattice proteins considered are plotted against their relative contact order
(RCO) for thee values of the solvation energy parameter: $\varepsilon_s=0$
(filled squares), $-0.3\varepsilon$ (open circles) and $-0.6\varepsilon$
(filled circles), as indicated.}
\label{fig6}
\end{figure}

\begin{figure*}[!ht]
\includegraphics[width=5.in]{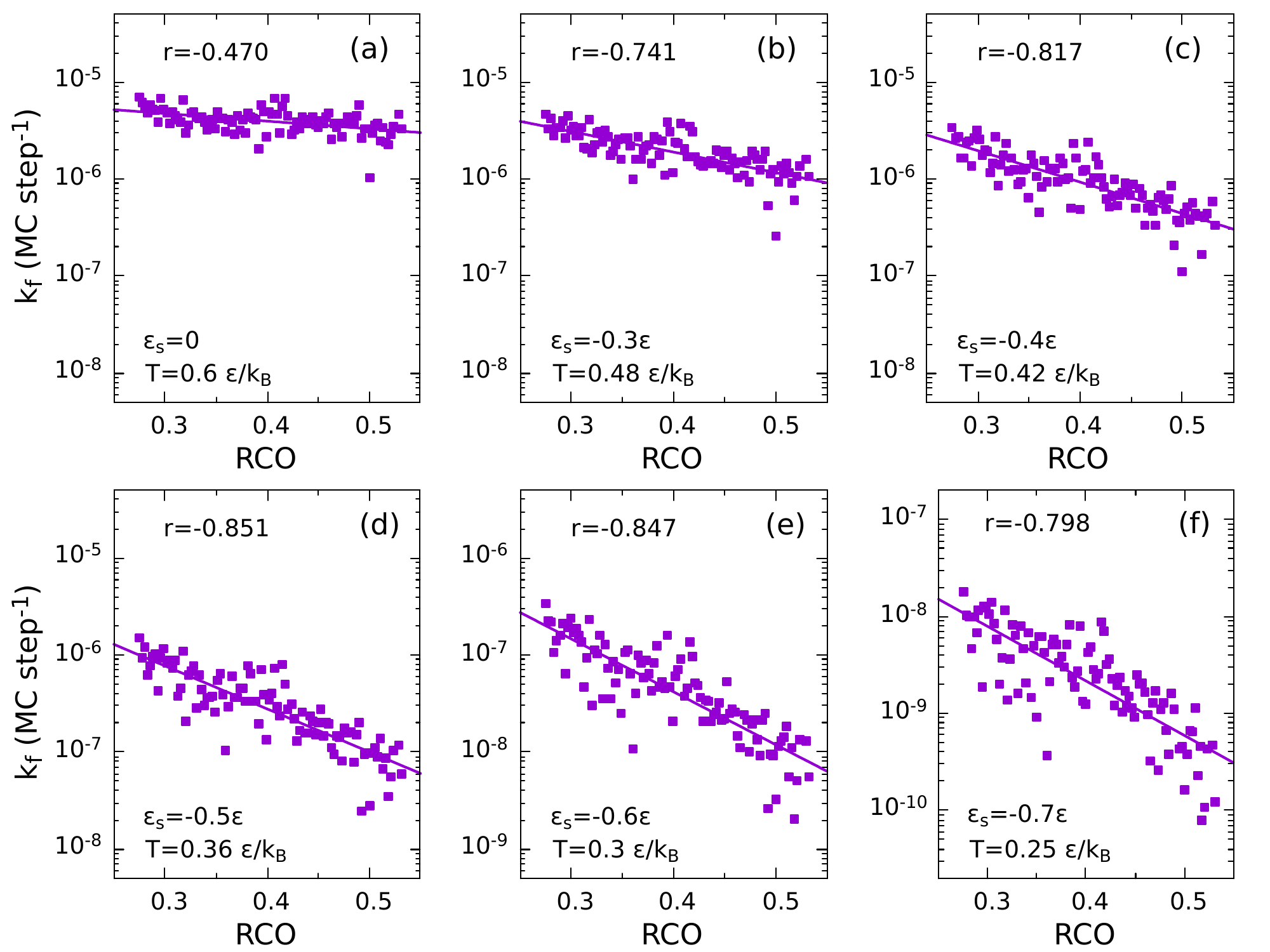}
\caption{Folding rate ($k_f$), on a logarithmic scale, vs. the relative contact
order (RCO) of 97 lattice proteins, as considered in Fig. \ref{fig6}, is shown
for 6 cases (a--f) with different values of $\varepsilon_s$ and $T$, as
indicated.  A linear fit to the data points (solid line) and the
correlation coefficient $r$ are provided in each plot. The corresponding
$p$-value is smaller than $10^{-5}$ in all cases. Each data point was
obtained from 201 independent folding trajectories.}
\label{fig7}
\end{figure*}

We now turn to examining the effect of solvation on the correlations between
folding properties and the relative contact order. To this end, we considered
97 model proteins with distinct RCO values and determined their specific heats
and folding rates through simulations.  Figure \ref{fig6}a shows that the
temperature $T_\mathrm{max}$, corresponding to the maximum of the specific heat
of the proteins, increases weakly with RCO, albeit with some small variations.
Without solvation, $T_\mathrm{max}$ rises by approximately 8\% as RCO increases
from 0.28 to 0.53. With solvation, the rise decreases to about 6\% for
$\varepsilon_s= -0.3\,\varepsilon$ and about 4\%
for $\varepsilon_s=-0.6\,\varepsilon$.  Fig. \ref{fig6}b shows that the
cooperativity index with a baseline subtraction, $\kappa_2^{(s)}$, on average,
increases with RCO.  Interestingly, the RCO dependence of this cooperativity
index is strongest for the case without solvation, $\varepsilon_s=0$, and
weakens as $\varepsilon_s$ decreases below zero. Similar trends are also found
for $\kappa_2$ and $\kappa_2^{(c)}$ (data not shown).
These results indicate that proteins with higher RCO in
general are thermally more stable and have higher folding cooperativity than
those with lower RCO. It is also shown that solvation weakens the RCO
dependence of both thermal stability and folding cooperativity, even though
it enhances folding cooperativity. These behaviors can be rationalized
as being due to the stabilizing effect of short-range native contacts (along
the sequence). In the absence of the solvation term, 
a larger number of such contacts are expected to be present in 
conformations of the unfolded state for proteins with lower RCO, leading to an
increased stability of the unfolded state and consequently a lower folding
temperature. More of these contacts also contribute to the stability of
partially folded conformations, leading to a decreased folding cooperativity
for proteins with lower RCO. In the presence of the solvation energy term,
the effect of short-range native contacts is diminished by
competing solvent interactions.

To examine the effect of solvation on the rate-topology dependence, we
determined the folding rates, $k_f$, at constant temperatures $T$'s below
$T_\mathrm{max}$ for all 97 proteins. We found that the best correlation
between $\log k_f$ and RCO occurs when $T$ is in the range from 
$0.85\, \overline{T}_\mathrm{max}$ to $0.9\, \overline{T}_\mathrm{max}$, 
where $\overline{T}_\mathrm{max}$ is
the mean value of $T_\mathrm{max}$ of all the proteins at a given
$\varepsilon_s$.  Figure \ref{fig7} shows the dependence of $\log k_f$ on RCO
at temperatures within this `optimal' range for various values of
$\varepsilon_s$. For each value of $\varepsilon_s$, this rate-topology
dependence is shown for a fixed temperature $T$ for all proteins to mimic
a real situation, as if the folding rates of all proteins were measured under
the same solvent conditions. In this figure, one observes not only the
tendency of $\log k_f$ to decrease with RCO in all the cases, but also the
increase the slope of the linear regression between
$\log k_f$ and RCO as $\varepsilon_s$ decreases.  
The diversity of the folding rates increases from less than one order of
magnitude for $\varepsilon_s=0$ to more than two orders of magnitude for
$\varepsilon_s=-0.7\varepsilon$. The correlation between $\log
k_f$ and RCO improves significantly as $\varepsilon_s$ decreases, with the
correlation coefficient $r$ rising from $-0.470$ to $-0.851$ as $\varepsilon_s$
decreases from 0 to $-0.5\varepsilon$ (Figs. \ref{fig7}a to \ref{fig7}d).  
However, the correlation somewhat decreases when $\varepsilon_s$ becomes
smaller than $-0.5\varepsilon$, but remains high at
$\varepsilon_s=-0.6\varepsilon$ and $-0.7\varepsilon$ (Figs. \ref{fig7}e and
\ref{fig7}f). The slope of the $\log k_f$ vs. RCO dependence, the
correlation coefficient $r$ and the $p$-value for all cases
studied with different values of $\varepsilon_s$ are listed in Table I.

\begin{table*}
\caption{Summary of the solvation dependence of folding cooperativity and the 
folding rate-topology correlation obtained from simulations for 97 lattice
proteins. For each value of the solvation energy parameter $\varepsilon_s$, the
data shown correspond to the mean values of the cooperativity indices
$\kappa_2$, $\kappa_2^{(c)}$, and $\kappa_2^{(s)}$, as well as the temperature
of the specific heat maximum, $T_\mathrm{max}$, averaged over all 97 proteins
(the average is denoted by a bar above each symbol).
Additionally, the temperature $T$ at which the folding rates $k_f$ of the
proteins were determined, the slope of the dependence of $\log k_f$ on RCO, 
the Pearson's correlation coefficient $r$, and the corresponding
$p$-value are provided.}
\begin{tabular}{ccccccccr}
\hline
$\varepsilon_s\,(\varepsilon)$ & 
~~~$\overline{\kappa}_2$~~~ & 
~~~$\overline{\kappa}_2^{(c)}$~~~ &
~~~$\overline{\kappa}_2^{(s)}$~~~ &
$\overline{T}_\mathrm{max} \,(\varepsilon/k_B)$ &
$T\,(\varepsilon/k_B)$ & Slope of $\log k_f$ vs. RCO & ~~~~~~$r$~~~~~~ & $p$-value\\
\hline
$ 0.0$ & 0.802 & 0.826 & 0.890 & 0.702 & 0.60 & $-0.79 \pm 0.15$ & $-0.470$ & $<10^{-5}$ \\
$-0.1$ & 0.839 & 0.858 & 0.908 & 0.658 & 0.57 & $-0.98 \pm 0.16$ & $-0.521$ & $<10^{-5}$ \\
$-0.2$ & 0.866 & 0.886 & 0.929 & 0.609 & 0.53 & $-1.26 \pm 0.18$ & $-0.628$ & $<10^{-5}$ \\
$-0.3$ & 0.888 & 0.912 & 0.952 & 0.553 & 0.48 & $-2.11 \pm 0.20$ & $-0.741$ & $<10^{-5}$ \\
$-0.4$ & 0.887 & 0.923 & 0.969 & 0.490 & 0.42 & $-3.24 \pm 0.23$ & $-0.818$ & $<10^{-5}$ \\
$-0.5$ & 0.875 & 0.929 & 0.983 & 0.423 & 0.36 & $-4.44 \pm 0.28$ & $-0.851$ & $<10^{-5}$ \\
$-0.6$ & 0.868 & 0.949 & 0.993 & 0.355 & 0.30 & $-5.46 \pm 0.35$ & $-0.847$ & $<10^{-5}$ \\
$-0.7$ & 0.850 & 0.950 & 0.995 & 0.295 & 0.25 & $-5.62 \pm 0.44$ & $-0.798$ & $<10^{-5}$ \\
\hline
\end{tabular}
\label{tab1}
\end{table*}

The best correlation obtained in the present study ($r \approx -0.85$) is
better than that in the study of Jewett et al. \cite{Plaxco2003} ($r\approx
-0.75$), but somewhat lower than that in the work of Kaya and Chan
\cite{Kaya2003prot} ($r\approx -0.91$), for similar lattice systems but with
different models. The increase in the diversity of the folding rates due to
the effect of solvation in our study is not too impressive but comparable to
those in the previous studies \cite{Plaxco2003,Kaya2003prot}. 

We have also calculated the average values of various cooperativity
indices, $\overline{\kappa}_2$, $\overline{\kappa}_2^{(c)}$,
and $\overline{\kappa}_2^{(s)}$,
across all 97 proteins, for different values of
$\varepsilon_s$ (see Table I). The dependences of these average indices on 
$\varepsilon_s$ are similar to that shown in Fig. \ref{fig2}(b). 
We found that the correlation coefficient $r$ as well as the slope
of the $\log k_f$ vs.  RCO dependence strongly correlate with 
both $\overline{\kappa}_2^{(c)}$ and $\overline{\kappa}_2^{(s)}$,
whereas it shows a medium correlation with $\overline{\kappa}_2$.
In particular, the corresponding correlation coefficient exceeds 0.96 for 
$r$ vs. $\overline{\kappa}_2^{(c)}$ and $r$ vs. $\overline{\kappa}_2^{(s)}$,
and is about 0.78 for $r$ vs. $\overline{\kappa}_2$.
Thus, the non-monotonic trends observed in the dependencies of
$\overline{\kappa}_2$ and $r$ on $\varepsilon_s$ seem to be weakly related. The
strong correlations of $r$ with $\overline{\kappa}_2^{(c)}$
and $\overline{\kappa}_2^{(s)}$ support the idea that the
folding rate-topology dependence is linked to folding cooperativity
\cite{Plaxco2003}.

\section{Discussion}

Although a more appropriate way to theoretically study protein folding
would be to start with a sequence-based model and select foldable sequences
through an evolutionary sequence design \cite{Shakh2004,Faisca2002}, in this
work, we adhere to the G\=o model due to the significant success of its
structure-based
approach in capturing protein folding mechanisms  \cite{Baker2000} and its
relatively well-known folding properties \cite{Cieplak_prl}. Because G\=o model
is minimally frustrated \cite{Bryngelson1987}, its energy landscape is smooth
and funnel-like \cite{Nymeyer1998}, facilitating rapid folding, as expected for
globular proteins.  The folding rate in the G\=o model is close to that of the
fastest-folding sequence in a comparable sequence-based model for a given
native structure \cite{Plotkin2001,Cieplak2002}. G\=o model is also often more
cooperative than its sequence-based counterparts \cite{Kaya2000prot}. Using the
G\=o model, the solvation effects on folding cooperativity and the rate-topology
dependence can be studied without perturbations from the sequence
\cite{Plaxco2000}.  For the type of solvation energy considered, the solvation
effects can be expected to be independent of the sequence.  Thus, for
sequence-based models, the effects of solvation can be qualitatively similar to
those obtained with the G\=o model.

To some extent, the favorable solvation energy for fully exposed
residues in our model may correspond to the solvation of peptide groups
(-NHCO-) in the protein backbone.  Experimental transfer data of amides
indicate that the polar interaction between water and peptide CO and
NH groups is entirely enthalpic with the hydration enthalpy as low as
$-11.6$~kcal/mol for a free amide group \cite{Baldwin1999}.
Electrostatic solvation free energy (ESF) calculation based on numerically
solving the Poisson-Boltzmann equation \cite{Honig1994} gives an ESF of
$-7.9$~kcal/mol for an alanine peptide group in the solvent-exposed
$\beta$-strand conformation, and $-2.5$~kcal/mol for a hydrogen-bonded
(H-bonded) peptide group in a solvent-exposed alanine $\alpha$-helix
\cite{Baldwin2000}. A similar ESF of $-2.5$~kcal/mol was found also for an
alanine H-bonded peptide group in an
alanine $\beta$-hairpin \cite{Baldwin2002,Baldwin2003}. Other non-polar amino
acids may have higher ESF values (negative but closer to zero) for their
peptide groups in $\alpha$- and $\beta$-structures than alanine due to more
effective side-chain shielding of the peptide backbone \cite{Baldwin1999}.
These data indicate that the desolvation of a peptide group due to secondary
structure formation increases the hydration enthalpy by at least 5.4~kcal/mol.
This energy penalty is significantly large compared to an average contribution
of hydrophobic interaction, which has been estimated to be about
$-1.25$~kcal/mol per residue \cite{Baldwin2003}, based on the calculation of
buried nonpolar surface areas.  If the secondary structures form
prior to the tertiary structure \cite{Karplus1994,Hoangjcp2,Dill2003}, then for most residues, the
first contact with another residue would correspond to the formation of a
peptide H-bond, which causes a substantial desolvation of the peptide group. 
In our model, a peptide H-bond is considered as a native interaction.
A residue can undergo multiple stages of desolvation during the folding
process. The above consideration suggests that the initial stage of desolvating
the peptide group due to peptide H-bond formation is associated with the most
significant change in the hydration enthalpy. The singular term in our
solvation potential is supported by this assessment.

\begin{figure}
\includegraphics[width=3.4in]{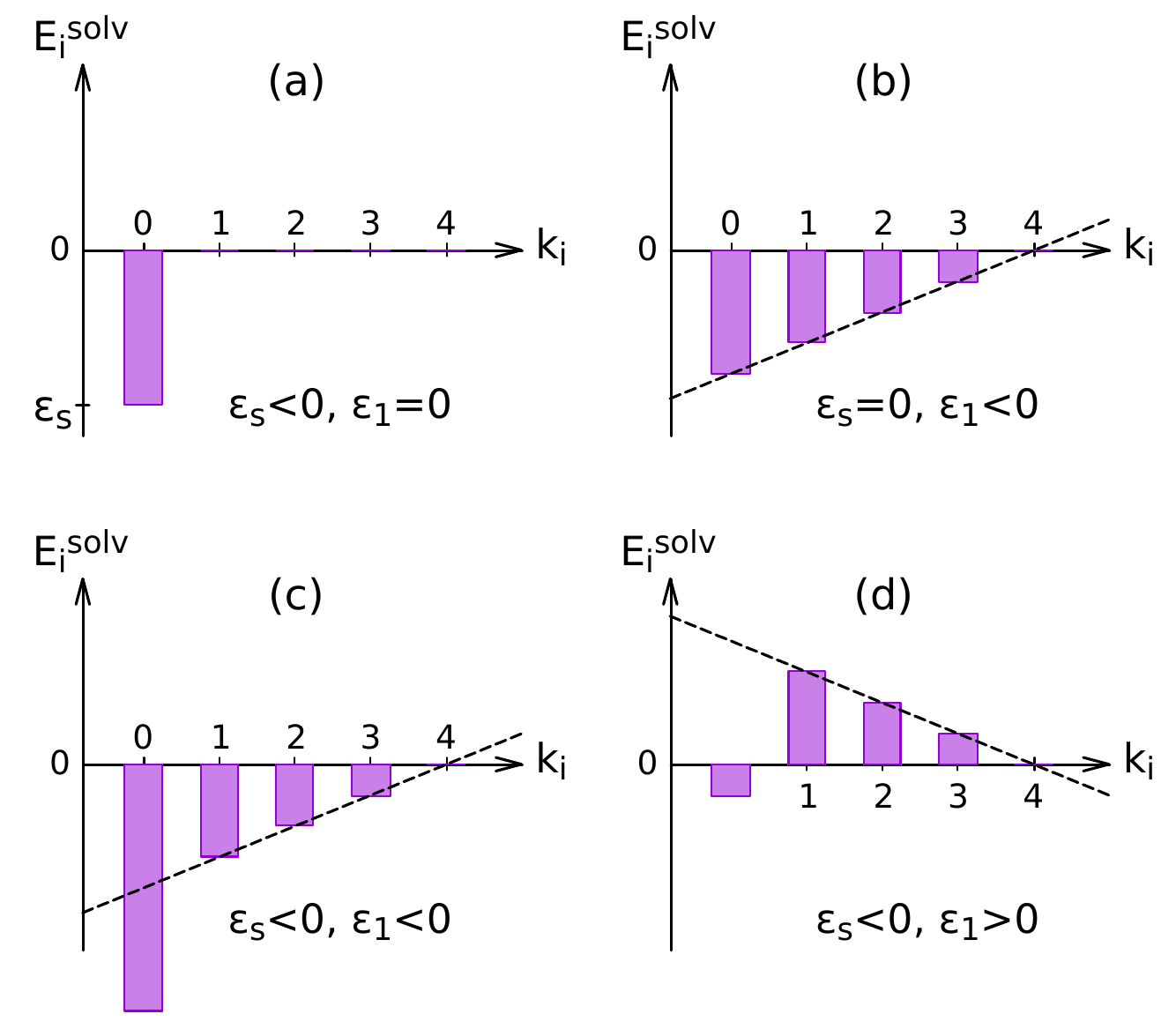}
\caption{Different shapes of the solvation potential, shown as dependence
of the residual solvation energy ($E_i^\mathrm{solv}$) on the number of
residue-residue contacts ($k_i$) of a non-terminal residue. $E_i^\mathrm{solv}$
is obtained from Eq. (\ref{eq5}) for different choices of parameters: 
(a) $\varepsilon_1=0$, $\varepsilon_s<0$; 
(b) $\varepsilon_1<0$, $\varepsilon_s=0$; 
(c) $\varepsilon_1<0$, $\varepsilon_s<0$; 
and (d) $\varepsilon_1>0$, $\varepsilon_s<0$; 
The energy unit is arbitrary.  The linear part of the potential is indicated by
a dashed line.}
\label{fig8}
\end{figure}

One can think of a more general form of the considered solvation
potential, which covers also partially exposed residues.  Let $k_i$ be the
number of contacts made by residue $i$ in a given conformation; $k_i$ takes
integer values from 0 to $k_i^\mathrm{max}$, where $k_i^\mathrm{max}$ is equal
to 4 and 5 for non-terminal and terminal residues, respectively, on the cubic
lattice. It
follows that $\sum_i k_i = 2 (n_c + n_{nn})$ and $\sum_i \delta_{k_i,0} = n_0$,
where $n_{nn}$ is the number of non-native contacts, and $\delta_{k,l}$ is the
Kronecker's delta function.  Consider the residual solvation energy as a
function of $k_i$ in the form:
\begin{equation}
E_i^\mathrm{solv} = (k_i^\mathrm{max} - k_i)\varepsilon_1 
+ \delta_{k_i,0} \, \varepsilon_s ,
\label{eq5}
\end{equation}
where 
$(k_i^\mathrm{max} - k_i)$ represents the exposure degree of the residue $i$,
and $\varepsilon_1$ is an energy parameter. Incorporating the solvation
potential in Eq.  (\ref{eq5}) into the G\=o model gives the total energy
\begin{equation}
E= -n_c(\varepsilon+2\varepsilon_1) - 2 n_{nn} \varepsilon_1 + n_0
\varepsilon_s  + \text{const.} \ , 
\label{eq6}
\end{equation}
where the additive constant does not depend on the chain conformation.
Thus, the linear term in Eq. (\ref{eq5}) is interchangeable with the energies
of native and non-native contacts, and therefore can be absorbed into the G\=o
model with energy contributions from non-native contacts.

Figure \ref{fig8} shows various shapes of the solvation potential in Eq.
(\ref{eq5}) for several choices of the parameters $\varepsilon_s$ and
$\varepsilon_1$.  We have checked that the solvation potential in Fig.
\ref{fig8}b ($\varepsilon_s=0$, $\varepsilon_1<0$) enhances the folding
cooperativity of the G\=o model in a similar manner to what obtained by the
potential in Fig. \ref{fig8}a ($\varepsilon_s<0$, $\varepsilon_1=0$), but to a
significantly lower degree. In particular, by using the same
analysis of the specific heats as shown in Fig. \ref{fig2}, as $\varepsilon_1$
decreases towards more negative values, the cooperativity index
$\kappa_2^{(s)}$ increases only up to a maximum of 0.91, while $\kappa_2$
reaches a maximum of 0.81. It is expected that the potential in Fig.
\ref{fig8}c ($\varepsilon_s<0$, $\varepsilon_1<0$) can improve the folding
cooperativity as effectively as the potential in Fig. \ref{fig8}a, while the
potential in Fig. \ref{fig8}d ($\varepsilon_s<0$, $\varepsilon_1>0$) may be
less effective than the latter.  This expectation is based on the observation
that the competition between solvation and native interactions promotes
cooperativity.

The derivation of Eq. (\ref{eq6}) is similar to the treatment of
polymer-solvent interactions in the Flory-Huggins theory \cite{Doi1996}. It
indicates that if the solvation potential is a linear function of the number of
residue-residue contacts then the solvation energy remains pairwise additive.
In our model, pairwise non-additivity \cite{Kollman1995} comes from a singular
term for the solvation of fully exposed residues, which makes the solvation
potential nonlinear.
In traditional implicit solvent models \cite{Caflish2002}, the solvation free
energy of a solute is often assumed to be proportional to the solvent
accessible surface area (SASA) \cite{Lee1971} of its molecular surface.
Depending on the solvent molecule radius, SASA can be a nonlinear function of
the number of residue-residue contacts due to the overlaps of buried areas
arising from different contacts \cite{Street1998}. The nonlinearity of a
solvation potential for a residue may also arise from the irregular shape of
the residue and the anisotropy of its surface \cite{Hao1997}. For example, the
surface of an amino acid residue can have both polar and nonpolar regions,
which contribute differently to the solvation energy.

Several solvation models have been proposed in the past to capture
the effects of water on protein folding.
Early simulations of lattice protein models by Hao and Scheraga \cite{Hao1997}
have shown that adding a solvation term to pairwise contact potentials improves
the foldability of sequences and makes the folding transition first-order like.
Their solvation model specifies a preferred solvation state for each residue
and applies an increasing energy penalty for solvation states that deviate from
the preferred state. Using the Hao and Scheraga's model and a two-letter
sequence design, Sorenson and Head-Gordon found that sequences in the solvation
model fold faster and more cooperative than sequences in the nonsolvation model
\cite{Sorenson1998}. Note that Hao and Scheraga's solvation potential is also
nonlinear, depending on the type of amino acid, but it has no singular term
as in our potential.
Using a coarse-grained model for protein structure prediction,
Wolynes and coworkers \cite{Papoian2004} demonstrated that adding
nonpairwise-additive water-mediated knowledge-based interactions to the
Hamiltonian markedly improves the quality of structure prediction. Their
solvation model includes a water-mediated second-well potential that depends on
the local density environment of residues and facilitates residue-residue
indirect contact at an intermediate C$_\beta$-C$_\beta$ distance
(6.5--9.5~{\AA}), mimicking the effect of one or two water layers between the
residues.  This potential provides a `wetting' of non-buried residues, but
unlike our solvation potential, it has a distance constraint on the 
interacting residues, therefore promoting intermediate chain
compaction.  The studies of Chan and coworkers
\cite{Chan2003jmb,Chan2005,Chan2008,Chan2009,Chan2011,Chan2013}
showed that G\=o-like models with desolvation barriers embedded in the native
contact potentials \cite{Cheung2002,Chan2005} possess higher degrees of
folding cooperativity than models without the desolvation barriers. These
barriers account for the process of the hydration water expulsion when residues
come into a direct contact \cite{Levy2006}. It was shown that their effects on
folding cooperativity can be very strong and may be intimately linked to
the shortening of the effective attractive range of residue-residue interactions
\cite{Chan2013}.
Note that both the second-well potential and the desolvation barrier potential
in these previous works promote solvent-induced interactions at an intermediate
to short range of distance between the residues, while our solvation energy
applies to fully exposed residues regardless of their distance from each other.
However, a favorable interaction with the solvent is implied in
both our and their models.
The successes of all these previous approaches demonstrate the significant role
of water in various details of the folding process \cite{Levy2004,Levy2006}.
Our study further supports this perspective.

Our study aligns with previous studies of lattice models, which showed
that higher folding cooperativity is associated with a stronger topology
dependence of the folding rates \cite{Plaxco2003,Kaya2003prot}. However,
other studies of off-lattice models indicated that while cooperativity
increases folding rate diversity, it does not necessarily enhance the
rate-topology dependence \cite{Chan2008,Chan2009,Chan2011,Chan2013}. 
The mechanism behind the folding rate-topology dependence may lie beyond the
energetics \cite{Debe1999,Eaton1999,Dill2003,Makarov2003,Chan2005topomer}.
Notably, a model based on the ``zipping up'' mechanism, using a few physical
rate parameters, can predict the folding rates of two-state proteins
from their native contact maps with good correlation to experimental data
\cite{Dill2003}.  This mechanism is defined by a time sequence of folding
events supported by molecular dynamics simulations of a G\=o-like model
\cite{Hoangjcp1,Hoangjcp2}.  
A number of studies have been focused on the topological aspect of the
rate-topology relation. In addition to RCO, several descriptors have
been proposed to predict folding rates from the native state structures, 
including long-range order \cite{Gromiha2001},
the number of native contacts \cite{Makarov2002},
total contact distance \cite{Zhou2002}, 
absolute contact order \cite{Finkelstein2003},
cliquishness \cite{Micheletti2003},
local secondary structure content \cite{Rose2003},
and relative logarithmic effective contact order \cite{Dixit2006}.
Although their performance typically does not surpass that of RCO for two-state
proteins, some are effective for both two- and three-state proteins
\cite{Zhou2002,Finkelstein2003,Micheletti2003}.
Recently, it was shown that maximum intrachain contact entanglement
\cite{Baiesi2017}, a new descriptor of protein native state entanglement, can
improve the folding rate predictions when combined with other descriptors,
underscoring the importance of topological complexity beyond contact
order.

\section{Conclusion}

We have introduced a solvation energy into the lattice G\=o model, which
favors residues that are fully exposed to the solvent. This solvation energy
promotes unfolding of proteins and is meant to compensate for the
overwhelming native bias in the G\=o model which considers only native
interactions. Our results show that it significantly enhances both the folding
cooperativity and the folding rate-topology dependence of the lattice proteins
considered, thereby suggesting that this type of solvent interactions may play
a key role in determining these properties in two-state proteins.

The mechanism by which the competition between the solvation and
native interactions enhances folding cooperativity has been elucidated. We
have demonstrated that as the solvation energy parameter decreases, the folding
free energy barrier increases. This occurs due to a decrease in the free energy
of the unfolded state together with an increase in the free energy of the
transition state. In addition, the folding pathways become narrower because
the solvation potential energetically disfavors the formation of non-native
contacts. The heightened folding free energy barrier and the narrowed folding
routes strengthens folding cooperativity by making the system more
two-state-like. 

We have also suggested that the solvation energy in our model 
corresponds to the solvation of peptide groups in the protein backbone. 
The polar interaction between water and exposed peptide groups is highly
favorable. Peptide groups
become substantially desolvated due to peptide hydrogen bonding during
secondary structure formation, resulting in a substantial deficit in hydration
enthalpy. Our study indicates that the effects of peptide group solvation
can be significant for the protein folding cooperativity and for the
folding rate-topology dependence. It is expected that these effects can be
studied in more realistic models.

\section*{Acknowledgements}

This work is dedicated to Marek Cieplak and is supported by Vietnam Academy of
Science and Technology under Grant No. NCXS02.05/22-23. 
The simulations were conducted using the HPC cluster at IOP-VAST.

\section*{Author Declarations}

\subsection*{Conflict of interest}

The authors have no conflicts to disclose.

\section*{Data Availability Statement}

The data that support the findings of this study are available from
the corresponding author upon reasonable request.

\bibliography{refs_solv}
\bibstyle{apsrev}

\end{document}